# Baroclinic Annular Mode Modulates Global Atmospheric Rivers and Extreme Precipitation

**Yuan-Bing Zhao[1], Lei Wang[1], and Yi Ming[2]**

[1] Department of Earth, Atmospheric and Planetary Sciences, Purdue University; West Lafayette, Indiana 47907, USA.

[2] Schiller Institute for Integrated Science and Society Department of Earth and Environmental Sciences, Boston College; Chestnut Hill, MA 02467.

Corresponding author: Yuan-Bing Zhao (zhao1550@purdue.edu); Lei Wang (leiwang@purdue.edu)

**Key Points:**

- Baroclinic annular mode systematically reorganizes atmospheric river (AR) corridors and extreme precipitation across both hemispheres.
- BAM modulation of AR and extreme precipitation is more zonally coherent in the SH and more regionally structured in the NH.
- BAM modulation of AR and extreme-precipitation occurrence locally exceeds 50% across major extratropical storm-track regions.

## Abstract

The Baroclinic Annular Mode (BAM) strongly influences extratropical circulation, yet its role in the global hydrological cycle remains unclear. Using ERA5 reanalysis and a global atmospheric river (AR) database, we show that BAM systematically reorganizes AR occurrence and extreme precipitation in both hemispheres. During high-BAM states, AR frequency increases along the flanks of major storm tracks and decreases near their climatological cores, redistributing preferred AR corridors. Relative AR-frequency changes locally exceed 50%, with the most zonally coherent patterns over the Southern Ocean. Extreme precipitation varies correspondingly, with 99th-percentile precipitation occurrence increasing by more than 50% across broad poleward storm-track regions. A large fraction of these extremes coincide with ARs, and AR-associated extreme precipitation shows similar BAM-related modulations. Low-BAM states show broadly opposite anomalies. These results identify BAM as a key mode linking large-scale circulation to global moisture transport and hydrological extremes.

## Plain Language Summary

Atmospheric rivers transport large amounts of water vapor and are responsible for many of the world's heaviest precipitation events. We show that a large-scale pattern of atmospheric variability, known as the baroclinic annular mode (BAM), systematically shifts where atmospheric rivers occur around the globe. During high-BAM states, atmospheric rivers become more common on the flanks of storm tracks and less common near their climatological centers. These changes are closely mirrored by shifts in extreme precipitation, demonstrating that BAM plays an important role in regulating global atmospheric moisture transport and precipitation extremes.

## 1 Introduction

Day-to-day weather in the midlatitudes is shaped largely by eastward-propagating baroclinic cyclones and anticyclones. These transient disturbances are organized into storm tracks, where eddy kinetic energy is concentrated and the atmosphere transports heat, momentum, and moisture (Chang, Lee, & Swanson, 2002). Variations in storm-track intensity and structure therefore influence a broad range of societally important weather, including strong winds, cloudiness, precipitation, and persistent circulation regimes (Shaw et al., 2016; Yau & Chang, 2020). The baroclinic annular mode (BAM) describes a dominant hemispheric-scale fluctuation in extratropical eddy activity, characterized by coherent variations in eddy kinetic energy and poleward heat transport on approximately 25–30-day timescales (Thompson & Barnes, 2014; Thompson & Li, 2015). Previous studies have focused primarily on its dynamics, linking BAM variability to eddy heat and momentum fluxes, downstream development, transient wave packets, and atmospheric blocking (Thompson & Woodworth, 2014; Wang & Nakamura, 2015; Nakayama et al., 2021, 2023; Liu & Wang, 2024). BAM-related variations in precipitation, clouds, and cloud-radiative effects have also been documented (Thompson & Li, 2015; Li & Thompson, 2016; Wang, Lu, & Kuang, 2018; Liu & Wang, 2023), suggesting that the influence of the BAM may extend beyond atmospheric dynamics to the extratropical hydrological cycle.

Baroclinic storms not only shape extratropical circulation but also organize much of the midlatitude moisture transport, providing a potential pathway through which the BAM may influence the hydrological cycle. Atmospheric rivers (ARs) are elongated corridors of intense water-vapor transport that play a central role in the extratropical hydrological cycle (Zhu & Newell, 1998; Guan & Waliser, 2015; Shields et al., 2018; Ong & Yang, 2024). ARs are closely coupled to extratropical cyclones and storm tracks, which organize the large-scale circulation and moisture transport associated with their development and propagation (Guo et al., 2020; Dacre & Clark, 2025). When ARs encounter frontal or orographic ascent, their sustained moisture transport can generate heavy precipitation, flooding, and strong winds (Lavers & Villarini, 2013; Waliser & Guan, 2017). At the same time, ARs provide an important fraction of annual precipitation and freshwater supply in many regions (Paltan et al., 2017). Understanding variability in AR occurrence, intensity, and pathways is therefore important for understanding how large-scale circulation variability translates into changes in water availability and precipitation extremes.

Among the hydrological phenomena closely coupled to extratropical circulation, ARs provide an important link between leading modes of atmospheric variability and precipitation extremes. The barotropic annular modes redistribute AR activity between subpolar and subtropical latitudes and produce corresponding changes in precipitation (Baek, Battalio, & Lora, 2023). North Atlantic Oscillation (NAO)-related jet shifts alter European AR landfall locations (Zavadoff & Kirtman, 2020). On longer timescales, the Interdecadal Pacific Oscillation (IPO) and Atlantic Multidecadal Oscillation (AMO) influence interdecadal Arctic AR trends (Ma et al., 2024). Wang et al. (2020) demonstrated that low-frequency, large-scale circulation changes in the Arctic play a crucial role in regulating AR activity. Tropical variability also affects extratropical ARs: the Madden–Julian Oscillation (MJO) and the El Niño–Southern Oscillation (ENSO) modulate North Pacific AR activity and landfall through their influence on large-scale circulation and tropical–extratropical teleconnections (Guan et al., 2012; Mundhenk, Barnes, & Maloney, 2016; Zhou, Kim, & Waliser, 2021). In the Southern Hemisphere, zonal wavenumber 3 has been associated with the orientation of flood-producing ARs near New Zealand (Kingston, Lavers, & Hannah, 2022). Collectively, these studies demonstrate that AR activity is closely tied to variability in the large-scale atmospheric circulation.

Despite these connections, the hydrological signature of the BAM on the global scale remains poorly understood. It remains unclear whether the hemispheric organization of eddy activity associated with the BAM is reflected in global AR occurrence, and whether variations in AR occurrence coincide with changes in precipitation extremes. Another open question is whether BAM variability produces an overall increase or decrease in these hydrological phenomena, or instead redistributes their occurrence across the midlatitudes.

Here, we examine these questions across both hemispheres and seasons. We use ERA5 (Hersbach et al., 2020) and a global AR database (Shields et al., 2018) to investigate the global hydrological signature of the Northern and Southern Hemisphere BAMs. We derive daily BAM indices from zonal-mean eddy kinetic energy over 1940–2024 and evaluate their hemispheric structures and intraseasonal variability. As in Liu & Wang (2024), we then contrast high- and low-BAM states separately for winter and summer in both hemispheres. Our analysis focuses on two complementary hydrological indicators: AR occurrence and the occurrence of daily 99th-percentile precipitation extremes. We show that BAM variability systematically reorganizes preferred AR corridors across seasons and hemispheres, with closely aligned variations in extreme precipitation. These results identify the BAM as an important mode linking hemispheric-scale baroclinic eddy variability to atmospheric river activity and precipitation extremes. The remainder of the paper is organized as follows: Section 2 and 3 describe the data and methods, Section 4 presents the results, and Section 5 provides a summary and conclusions.

**2 Methods**

The BAM indices are calculated following the eddy-kinetic-energy (EKE) framework of Thompson and Woodworth (2014) and Thompson and Li (2015), with hemisphere-specific zonal filtering. For each day, pressure level, and latitude, zonal wavenumbers 0-3 are removed for the Northern Hemisphere (NH), and zonal mean (i.e., wavenumber zero) is removed for the Southern Hemisphere (SH). Using the filtered winds $u^*$ and $v^*$, EKE is calculated as $\mathrm{EKE} = ½\,[u^{*2} + v^{*2}]$. At each pressure–latitude grid point, we remove the climatological seasonal cycle, defined as the annual mean plus the first four Fourier harmonics, from the daily zonal-mean EKE. February 29 is excluded, yielding 31,025 daily samples during 1940–2024. The resulting anomalies are mass- and area-weighted before empirical orthogonal function (EOF) analysis is performed over 200–1000 hPa and 20°–70° latitude separately in each hemisphere. The leading EOF mode defines the BAM, and its principal component is standardized to unit variance and oriented such that positive values correspond to enhanced EKE. Power spectra of the standardized daily BAM indices are estimated from 500-day Hanning-windowed segments with an overlap of 250 days. The segment periodograms are averaged at each positive frequency and smoothed once using three-point weights of 0.25, 0.50, and 0.25.

Following the methodology established in our recent work (Liu & Wang, 2024), we perform BAM-state composite analyses separately for each hemisphere and season: December–January–February (DJF) and June–July–August (JJA). DJF (JJA) corresponds to winter (summer) in the NH and summer (winter) in the SH. High- and low-BAM states are defined by standardized indices above +1 and below −1, respectively. Composite anomalies are reported relative to the corresponding all-day seasonal climatology.

Extreme precipitation is defined at each grid point as daily precipitation exceeding the local season-specific (DJF or JJA) 99th-percentile threshold during 1940–2024, with leap days excluded. By construction, extremes occur on 1% of days in each season at each grid.

## 3 Data

Daily-mean winds and total precipitation are obtained from ERA5 (Hersbach et al., 2020) and used to derive the BAM indices and identify precipitation extremes, respectively. All fields are regridded to a 1° × 1° latitude–longitude grid.

The AR occurrence is obtained from the ERA5-based hourly AR catalogs developed through the Atmospheric River Tracking Method Intercomparison Project (ARTMIP; Shields et al., 2018; Collow et al., 2022). The gridded catalog data have a spatial resolution of 0.25° × 0.25°. Eight global AR detection products are available. Our main analysis uses the Guan and Waliser product because it provides the longest record (1979–2019), near-global spatial coverage, and a well-established detection framework widely used in global AR research (Guan & Waliser, 2015; Waliser & Guan, 2017). Our main results are robust across all AR products examined; for brevity, three additional products showing representative, qualitatively consistent patterns are presented in the Supporting Information (Figures S1–S3). The original AR data consist of hourly binary grid-point tags indicating the presence or absence of an AR. We classify a grid point as experiencing an AR on a given day if at least one of its 24 hourly tags is nonzero. Thus, days with one and 24 tagged hours are treated equally, and multiple AR objects or episodes within the same day are not counted separately. For each season and BAM state, AR occurrence frequency is calculated as the number of AR-occurrence days divided by the number of valid days at each grid point. This metric therefore represents the fraction of days with AR occurrence, rather than the fraction of AR-tagged hours or the number of individual AR objects.

## 4 Results

### 4.1 Spatial Structure and Temporal Variability of the BAM

Figure 1 shows that the leading EOF of zonal-mean EKE exhibits a vertically coherent midlatitude structure in both hemispheres (panels a and c), consistent with the canonical BAM patterns reported by Thompson and Woodworth (2014) and Thompson and Li (2015). EOF1 explains 46.8% and 37.2% of the daily EKE variance in the NH and SH, respectively, with upper-tropospheric maxima near 45°N and 50°S. The corresponding principal components exhibit pronounced intraseasonal variability, with enhanced spectral power at periods of approximately 20–30 days (Figures 1b and 1d). Midlatitude-mean precipitation shows enhanced spectral power over a similar period range in both hemispheres, consistent with the previously documented relationship between BAM variability and midlatitude precipitation (Thompson & Li, 2015). Overall, these results show that the BAM indices used here reproduce the established spatial structure and intraseasonal variability of the BAM.

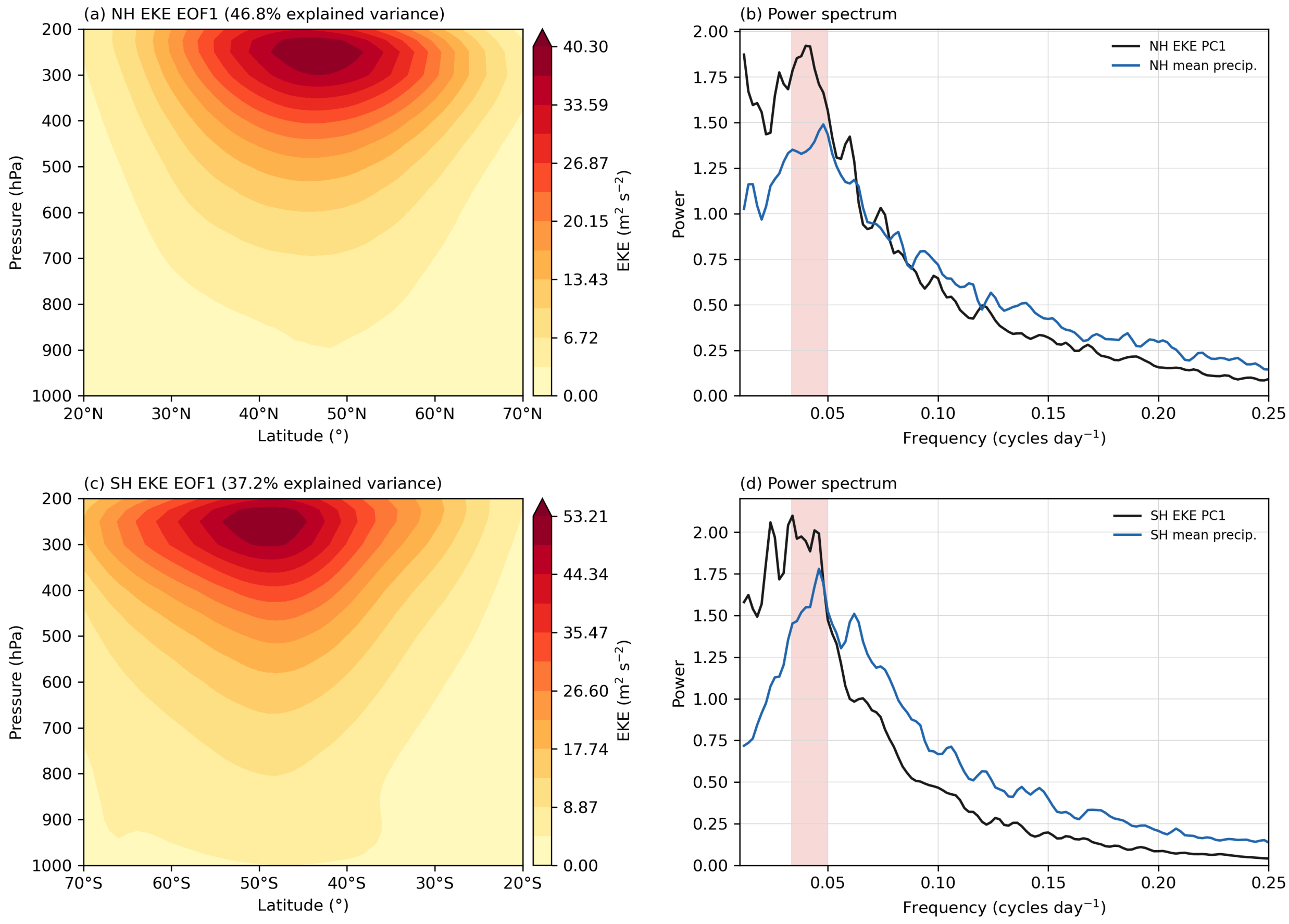

**Figure 1. Spatial and temporal characteristics of the BAM.** (a, c) Spatial structures of the leading mode, obtained by regressing zonal-mean EKE onto the standardized leading principal component. (b, d) Power spectra of the standardized leading principal components (black) and mean precipitation over the 20°–70° latitude band (blue). Red stripe denotes periods of 20–30 days. The top panels show the NH and the bottom panels show the SH.

### 4.2 BAM Modulation of AR Occurrence

Figure 2 presents BAM-state composite maps of AR occurrence combining winter (NH DJF and SH JJA) and summer (NH JJA and SH DJF) during 1979–2019. Shading shows the relative change in AR occurrence frequency, calculated as the difference between the BAM-state composite and the corresponding seasonal climatology divided by that climatology. Thick green contours denote the climatological AR corridors. Overall, BAM variability is associated with a large-scale reorganization of AR occurrence across the midlatitudes.

During winter, high-BAM states generally increase AR occurrence on the poleward and equatorward flanks of the climatological AR corridor while decreasing occurrence near their cores (Figure 2a). In the NH, the relative changes approach 75% over the coastal regions of East Asia and North America, and portions of the subtropical oceans, whereas moderate decreases near the climatological AR maxima over the North Pacific and North Atlantic. In the SH, the relative change in AR occurrence is more zonally coherent, with increases of up to approximately 50% at 50°–70°S, weak decreases at 35°–55°S, and strong increases farther equatorward. Low-BAM states largely reverse these patterns (Figure 2b). In the NH, AR occurrence decreases by up to approximately 50% along the poleward and equatorward flanks and increases slightly near the AR corridor cores. Over the Southern Ocean, AR occurrence decreases at 55°–70°S but increases by up to approximately 25% at 35°–55°S, particularly over the South Pacific. Together, these patterns point to a meridional redistribution of AR occurrence rather than a spatially uniform change in frequency.

During summer, BAM modulation of AR occurrence exhibits broadly similar patterns (Figures 2c and 2d). During high-BAM states, AR occurrence generally increases along the corridor flanks--by up to approximately 75% in the NH and 50% in the SH--and decreases near the corridor cores. Low-BAM states produce widespread reductions at higher midlatitudes, locally reaching approximately 75% and 50% in the NH and SH, respectively, together with localized increases on the equatorward side of the corridor (Figure 2d). Near 30°N, however, AR occurrence increases during both high- and low-BAM states over the Mediterranean and portions of the North Pacific and North Atlantic. The origin of this same-signed pattern remains unclear. In the SH, the summer response is more zonally uniform than the winter response. This zonal coherence, particularly over the Southern Ocean, is consistent with Park, Son, and Guan (2023), who showed that the zonally symmetric component of climatological AR activity is dominated by high-frequency moisture transport associated with storm tracks. The corresponding redistribution of AR occurrence between high- and low-BAM states suggests that this relationship also varies with fluctuations in storm-track eddy activity.

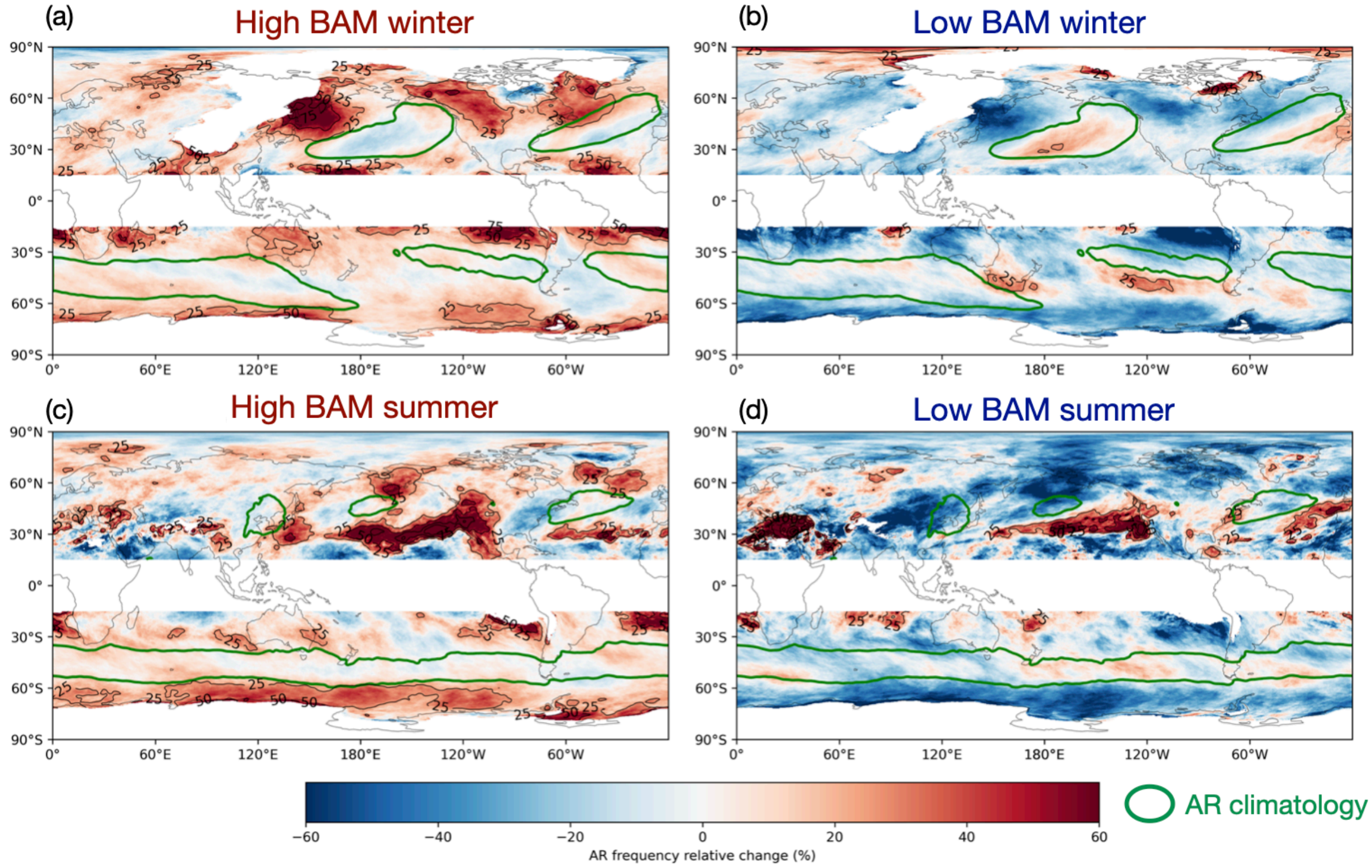

**Figure 2. BAM modulation of AR occurrence across hemispheres and seasons.** BAM-state composite of relative changes in AR occurrence for (a) high-BAM winter, (b) low-BAM winter, (c) high-BAM summer, and (d) low-BAM summer states. Winter and summer are defined locally: DJF and JJA, respectively, in the NH, and JJA and DJF, respectively, in the SH. Shading shows the difference between composite and climatological AR occurrence frequencies normalized by the corresponding seasonal climatology; black contours are plotted at 25% intervals. Thick green contours denote a climatological AR occurrence frequency of 30% for the corresponding season. The tropical region from 15°S to 15°N and grid points with climatology AR occurrence frequencies below 2% are masked.

### 4.3 BAM Modulation of Extreme Precipitation Occurrence

Because the intense moisture transport associated with ARs is closely linked to precipitation extremes, we next examine whether BAM modulation extends from AR occurrence to extreme precipitation. Figure 3 shows that BAM variability substantially modulates the occurrence of the 99th-percentile daily precipitation extremes across both hemispheres and seasons. During winter, high-BAM states are characterized by widespread increases in extreme precipitation, particularly across the poleward midlatitudes (Figure 3a). In the NH, the largest and most spatially coherent increases occur over the North Pacific sector, with relative changes exceeding 50% over broad regions and locally approaching or exceeding 100% around 60°N. In the SH, enhanced extreme precipitation forms a broad, nearly circumpolar band at high latitudes, with particularly strong increases over the Southern Ocean. Low-BAM winter states show broadly opposite patterns, with widespread reductions in extreme precipitation across many of the same regions (Figure 3b). During summer, the BAM-related patterns are more spatially heterogeneous in the NH (Figures 3c and 3d). High-BAM states show enhanced extreme precipitation over several midlatitude regions, interspersed with areas of reduced occurrence, while low-BAM states are characterized by widespread reductions at higher latitudes. In the SH, the pattern remains more zonally coherent, with high-BAM states associated with enhanced extreme precipitation across much of the Southern Ocean and low-BAM states associated with widespread reductions hemispherically. Overall, high- and low-BAM states exhibit broadly opposite patterns of extreme precipitation occurrence, with the clearest hemispheric-scale organization over the Southern Ocean.

Having established that BAM variability is associated with substantial changes in total extreme precipitation in Figure 3, we next examine the extent to which these extremes coincide with AR occurrence. ARs accompany a large fraction of extreme-precipitation days during both high- and low-BAM states, particularly over the major extratropical storm-track regions. Over much of the North Pacific, North Atlantic, and Southern Ocean, more than 50% of extreme-precipitation days coincide with AR occurrence, as indicated by the unhatched regions in Figure 4; the full spatial distributions of these fractions are shown in Figure S4. Here, we refer to these coincident events as “AR-associated” extreme precipitation; this designation indicates spatial and temporal co-occurrence and does not necessarily imply that the precipitation extreme is caused by the AR. We therefore next examine the AR-associated component of the BAM-related changes in extreme precipitation.

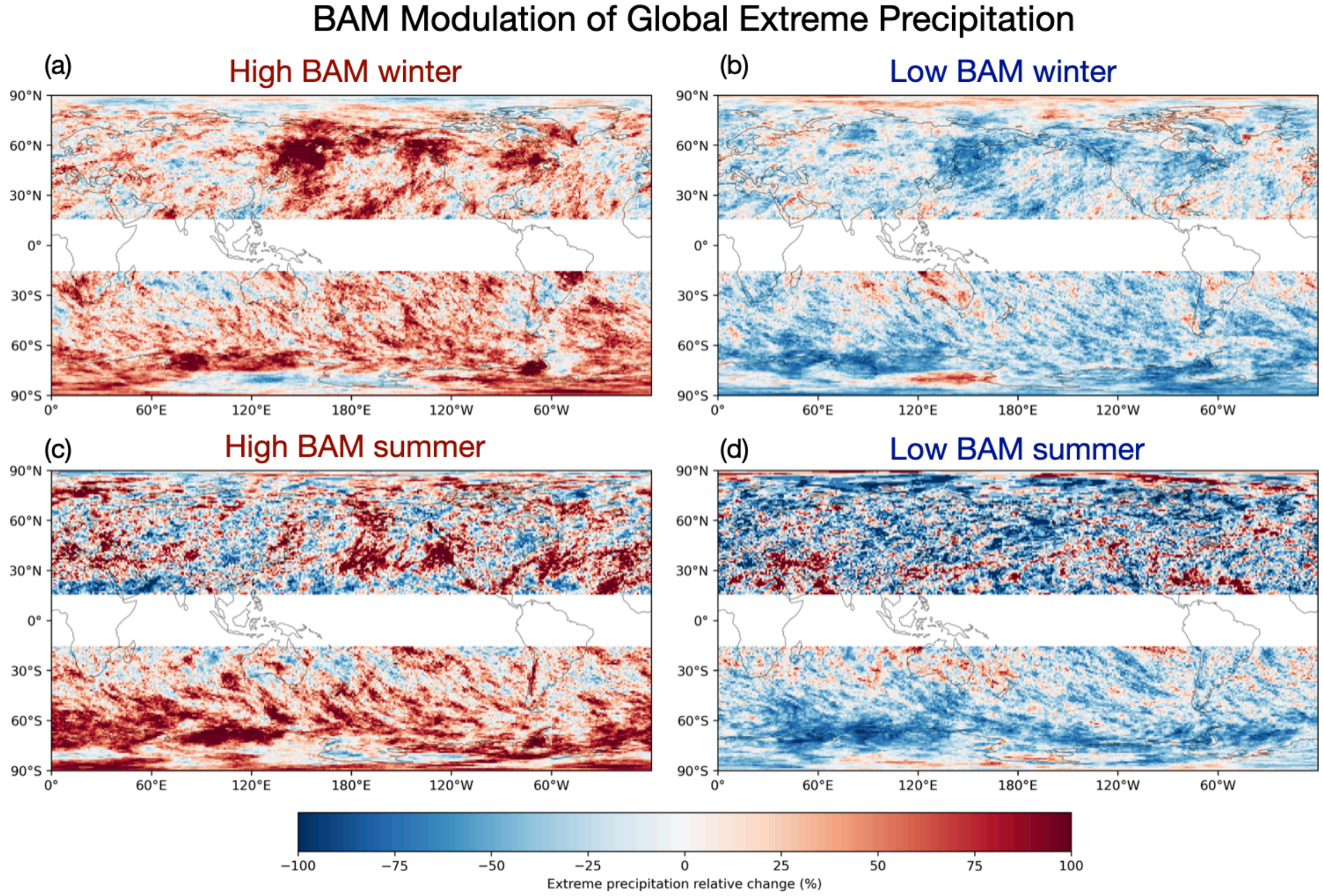

**Figure 3**. **BAM modulation of extreme precipitation across hemispheres and seasons.** As in Figure 2, but showing the relative changes in the occurrence of daily precipitation exceeding the local, season-specific 99th-percentile threshold during 1940-2024.

Figure 4 shows that AR-associated extreme precipitation exhibits pronounced and spatially coherent variations between high- and low-BAM states. AR-associated extreme precipitation is calculated at each grid point by multiplying the total relative change in extreme precipitation occurrence (Figure 3) by the fraction of extreme-precipitation days during each BAM state that coincide with a local AR. During winter, high-BAM states are associated with substantial increases over the North Pacific and North Atlantic, exceeding 50% across broad regions and locally approaching or exceeding 100% (Figure 4a). Particularly strong increases occur near the coasts of East Asia and North America, where AR occurrence also increases. In the SH, increases form a broad, nearly circumpolar band over the Southern Ocean near 55°–75°S, coinciding with the poleward enhancement of AR occurrence. Low-BAM states show broadly opposite patterns, with widespread reductions over the North Pacific and Southern Ocean and more regional changes over the North Atlantic (Figure 4b), where AR occurrence also decreases. During summer, the NH pattern is more regionally structured (Figures 4c and 4d). High-BAM states are associated with enhanced AR-associated extreme precipitation over several midlatitude regions, particularly over the North Pacific and North Atlantic, whereas low-BAM states show weaker and more spatially heterogeneous changes. In the SH, the pattern remains strongly zonally coherent, with widespread increases during high-BAM states and reductions during low-BAM states, particularly poleward of about 50°S. These patterns closely resemble the BAM-related changes in total extreme precipitation (Figure 3) and AR occurrence (Figure 2), particularly over the North Pacific, North Atlantic, and Southern Ocean. The spatial correspondence indicates that AR-associated events account for an important component of the BAM-related variations in extreme precipitation.

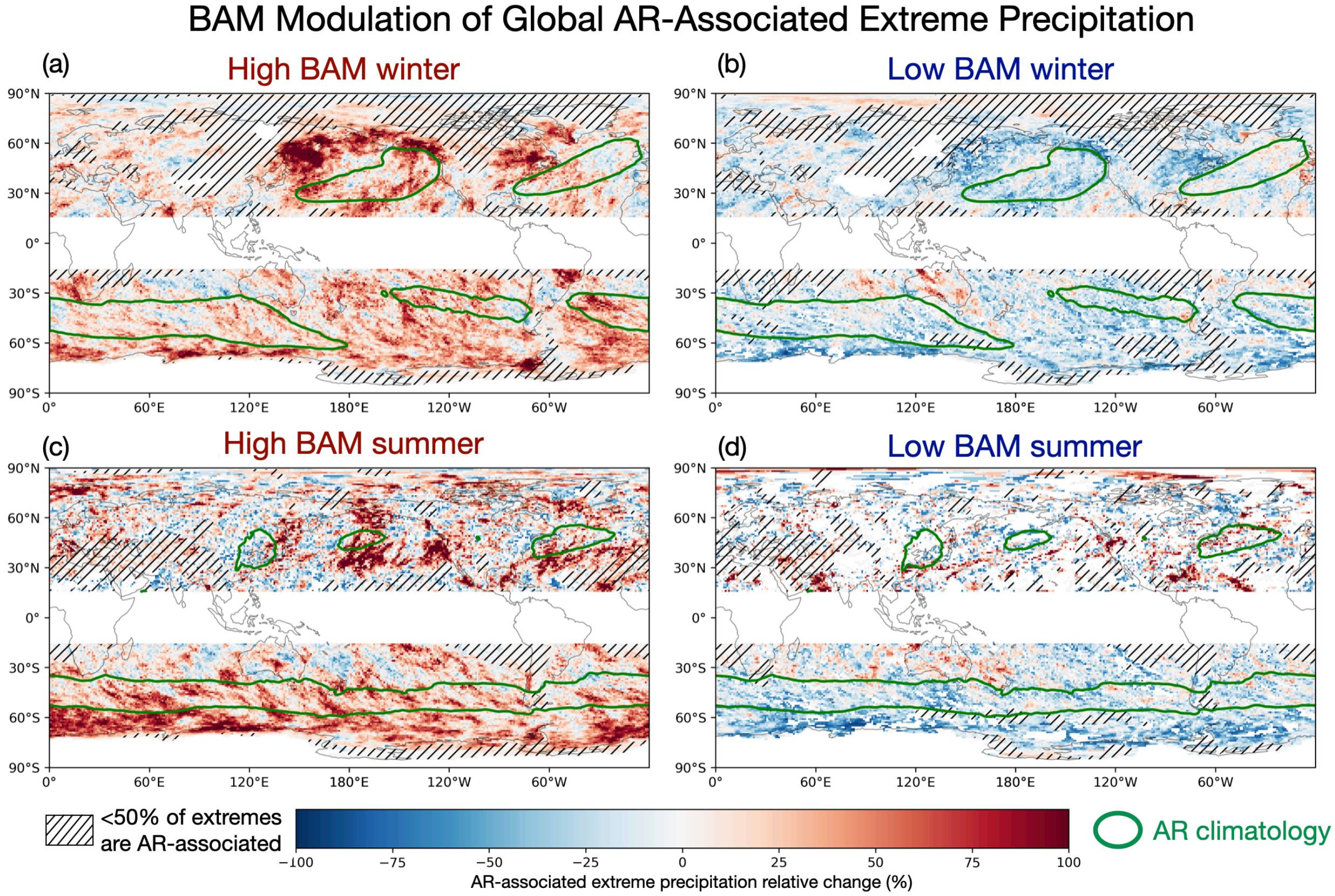

**Figure 4**. **BAM modulation of AR-associated extreme precipitation across hemispheres and seasons.** As in Figure 3, but showing the AR-associated BAM-related relative changes in the occurrence of daily precipitation exceeding the local, season-specific 99th-percentile threshold during 1940-2024. Thick green contours denote a climatological AR occurrence frequency of 30% for the corresponding season. Hatched regions indicate where less than 50% of extreme-precipitation days during the corresponding BAM state are associated with ARs.

## 5 Conclusions

BAM systematically modulates atmospheric river (AR) occurrence and extreme precipitation across both hemispheres and seasons. High-BAM states are associated with a meridional redistribution of AR occurrence, with enhanced occurrence along the flanks of climatological AR corridors and reduced occurrence near their cores. This redistribution is particularly zonally coherent in the SH, whereas the NH exhibits stronger regional structure. Low-BAM states show broadly opposite variations. Extreme precipitation varies correspondingly, with high-BAM states favoring more frequent extremes across many poleward midlatitude regions and low-BAM states generally associated with fewer extremes. The strongest and most spatially coherent variations occur over the major extratropical storm-track regions, particularly the North Pacific and Southern Ocean.

The close spatial correspondence among BAM-related variations in AR occurrence, AR-associated extreme precipitation, and total extreme precipitation indicates that ARs provide an important link between variations in hemispheric-scale baroclinic eddy activity and precipitation extremes. This connection is especially evident over the North Pacific, North Atlantic, and Southern Ocean, where ARs accompany a large fraction of extreme-precipitation days. At the same time, differences between total and AR-associated extreme precipitation, particularly over land, point to contributions from non-AR precipitation systems and regional processes. Our results therefore extend the hydrological relevance of the BAM beyond previously documented variations in mean precipitation, identifying it as a mode of intraseasonal variability associated with the organization of both atmospheric moisture transport and precipitation extremes.

## Acknowledgments

The research is funded by the NOAA award NA24OARX431C0054-T1-01.

## Availability Statement

ERA5 global atmospheric reanalysis winds and precipitation data are publicly available through the

Copernicus Climate Change Service (C3S) Climate Data Store at

https://cds.climate.copernicus.eu/datasets/reanalysis-era5-complete?tab=d_download

The Tier 2, AR dataset is available through website (Collow et al., 2022) at

https://gdex.ucar.edu/datasets/d651018/dataaccess/.

Codes are available in the open repository (Zhao, 2026) at https://zenodo.org/records/21847623.

## Conflict of Interest Disclosure

The authors declare there are no conflicts of interest for this manuscript.

## References

Baek, S. H., J. M. Battalio, and J. M. Lora (2023), Atmospheric river variability over the last millennium driven by annular modes, *AGU Adv.*, 4, e2022AV000834, https://doi.org/10.1029/2022AV000834.

Chang, E. K. M., S. Lee, and K. L. Swanson (2002), Storm track dynamics, J. Clim., 15(16), 2163–2183, https://doi.org/10.1175/1520-0442(2002)015%3C02163:STD%3E2.0.CO;2.

Collow, A. B. M., et al. (2022), An overview of ARTMIP's Tier 2 Reanalysis Intercomparison: Uncertainty in the detection of atmospheric rivers and their associated precipitation, *J. Geophys. Res. Atmos.*, 127, e2021JD036155, https://doi.org/10.1029/2021JD036155.

Dacre, H.F., Clark, P.A. (2025), A kinematic analysis of extratropical cyclones, warm conveyor belts and atmospheric rivers. *npj Clim Atmos Sci* **8**, 97, https://doi.org/10.1038/s41612-025-00942-z.

Guan, B., D. E. Waliser, N. P. Molotch, E. J. Fetzer, and P. J. Neiman (2012), Does the Madden–Julian Oscillation influence wintertime atmospheric rivers and snowpack in the Sierra Nevada?, *Mon. Weather Rev.*, 140(2), 325–342, https://doi.org/10.1175/MWR-D-11-00087.1.

Guan, B., and D. E. Waliser (2015), Detection of atmospheric rivers: Evaluation and application of an algorithm for global studies, *J. Geophys. Res. Atmos.*, 120(24), 12,514–12,535, https://doi.org/10.1002/2015JD024257.

Guo, Y., T. Shinoda, B. Guan, D. E. Waliser, and E. K. M. Chang (2020), Statistical Relationship between Atmospheric Rivers and Extratropical Cyclones and Anticyclones. *J. Climate*, **33**, 7817–7834, https://doi.org/10.1175/JCLI-D-19-0126.1.

Hersbach, H., et al. (2020), The ERA5 global reanalysis, *Q. J. R. Meteorol. Soc.*, 146, 1999–2049, https://doi.org/10.1002/qj.3803.

Kingston, D. G., D. A. Lavers, and D. M. Hannah (2022), Characteristics and large-scale drivers of atmospheric rivers associated with extreme floods in New Zealand, *Int. J. Climatol.*, 42(5), 3208–3224, https://doi.org/10.1002/joc.7415.

Lavers, D. A., and G. Villarini (2013), The nexus between atmospheric rivers and extreme precipitation across Europe, *Geophys. Res. Lett.*, 40(12), 3259–3264, https://doi.org/10.1002/grl.50636.

Li, Y., and D. W. J. Thompson (2016), Observed signatures of the barotropic and baroclinic annular modes in cloud vertical structure and cloud radiative effects, *J. Clim.*, 29(13), 4723–4740, https://doi.org/10.1175/JCLI-D-15-0692.1.

Liu, Z., &  Wang, L. (2023), Regional features of the 20–30 Day periodic behavior in the Southern Hemisphere summer circulation. *Geophysical Research Letters*,  50, e2023GL104256, https://doi.org/10.1029/2023GL104256.

Liu, Z., and L. Wang (2024), Enhanced occurrence of atmospheric blocking in the Southern Hemisphere by baroclinic annular mode, *Geophys. Res. Lett.*, 51(4), e2023GL107343, https://doi.org/10.1029/2023GL107343.

Ma, W., et al. (2024), The role of interdecadal climate oscillations in driving Arctic atmospheric river trends, *Nat. Commun.*, 15, 2135, https://doi.org/10.1038/s41467-024-45159-5.

Mundhenk, B. D., E. A. Barnes, and E. D. Maloney (2016), All-season climatology and variability of atmospheric river frequencies over the North Pacific, *J. Clim.*, 29(13), 4885–4903, https://doi.org/10.1175/JCLI-D-15-0655.1.

Nakayama, M., H. Nakamura, and F. Ogawa (2021), Impacts of a midlatitude oceanic frontal zone for the baroclinic annular mode in the Southern Hemisphere, *J. Clim.*, 34(18), 7389–7408, https://doi.org/10.1175/JCLI-D-20-0359.1.

Nakayama, M., H. Nakamura, S. Okajima, and F. Ogawa (2023), Modulations of storm-track activity associated with the baroclinic annular mode, *J. Clim.*, 36(12), 4219–4234, https://doi.org/10.1175/JCLI-D-22-0377.1.

Ong, H., and D. Yang (2024), Vapor kinetic energy for the detection and understanding of atmospheric rivers, *Nat. Commun.*, 15, 9428, https://doi.org/10.1038/s41467-024-53369-0.

Paltan, H., D. Waliser, W. H. Lim, B. Guan, D. Yamazaki, R. Pant, and S. Dadson (2017), Global floods and water availability driven by atmospheric rivers, *Geophys. Res. Lett.*, 44, 10,387–10,395, https://doi.org/10.1002/2017GL074882.

Park, C., S.-W. Son, and B. Guan (2023), Multiscale nature of atmospheric rivers, *Geophys. Res. Lett.*, 50(10), e2023GL102784, https://doi.org/10.1029/2023GL102784.

Shaw, T. A., et al. (2016), Storm track processes and the opposing influences of climate change, Nat. Geosci., 9, 656–664, https://doi.org/10.1038/ngeo2783.

Shields, C. A., et al. (2018), Atmospheric River Tracking Method Intercomparison Project (ARTMIP): Project goals and experimental design, *Geosci. Model Dev.*, 11(6), 2455–2474, https://doi.org/10.5194/gmd-11-2455-2018.

Thompson, D. W. J., and E. A. Barnes (2014), Periodic variability in the large-scale Southern Hemisphere atmospheric circulation, *Science*, 343(6171), 641–645, https://doi.org/10.1126/science.1247660.

Thompson, D. W. J., and Y. Li (2015), Baroclinic and barotropic annular variability in the Northern Hemisphere, *J. Atmos. Sci.*, 72(3), 1117–1136, https://doi.org/10.1175/JAS-D-14-0104.1.

Thompson, D. W. J., and J. D. Woodworth (2014), Barotropic and baroclinic annular variability in the Southern Hemisphere, *J. Atmos. Sci.*, 71(4), 1480–1493, https://doi.org/10.1175/JAS-D-13-0185.1.

Waliser, D., and B. Guan (2017), Extreme winds and precipitation during landfall of atmospheric rivers, *Nat. Geosci.*, 10, 179–183, https://doi.org/10.1038/ngeo2894.

Wang, L., J. Lu, and Z. Kuang (2018), A robust increase of the intraseasonal periodic behavior of the precipitation and eddy kinetic energy in a warming climate, *Geophys. Res. Lett.*, 45(15), 7790–7799, https://doi.org/10.1029/2018GL078495.

Wang, L., and N. Nakamura (2015), Covariation of finite-amplitude wave activity and the zonal mean flow in the midlatitude troposphere: 1. Theory and application to the Southern Hemisphere summer, *Geophys. Res. Lett.*, 42, 8192–8200, https://doi.org/10.1002/2015GL065830.

Wang, Z., Ding, Q., Wu, R. *et al.* Role of atmospheric rivers in shaping long term Arctic moisture variability. *Nat Commun* **15**, 5505 (2024). https://doi.org/10.1038/s41467-024-49857-y.

Yau, A. M. W., and E. K. M. Chang (2020), Finding storm track activity metrics that are highly correlated with weather impacts. Part I: Frameworks for evaluation and accumulated track activity, J. Clim., 33(23), 10,169–10,186, https://doi.org/10.1175/JCLI-D-20-0393.1.

Zavadoff, B. L., and B. P. Kirtman (2020), Dynamic and thermodynamic modulators of European atmospheric rivers, *J. Clim.*, 33(10), 4167–4185, https://doi.org/10.1175/JCLI-D-19-0601.1.

Zhao, Y.-B. (2026), Data and codes for "Baroclinic Annular Mode Modulates Global Atmospheric Rivers and Extreme Precipitation" [Data set and software], Zenodo, https://doi.org/10.5281/zenodo.21847623.

Zhou, Y., H. Kim, and D. E. Waliser (2021), Atmospheric river lifecycle responses to the Madden–Julian Oscillation, *Geophys. Res. Lett.*, 48(3), e2020GL090983, https://doi.org/10.1029/2020GL090983.

Zhu, Y., and R. E. Newell (1998), A proposed algorithm for moisture fluxes from atmospheric rivers, *Mon. Weather Rev.*, 126(3), 725–735, https://doi.org/10.1175/1520-0493(1998)126<0725:APAFMF>2.0.CO;2

Supporting Information for

# Baroclinic Annular Mode Modulates Global Atmospheric Rivers and Extreme Precipitation

Yuan-Bing Zhao[1], Lei Wang[1], and Yi Ming[2]

1Purdue University, West Lafayette, IN USA.

2Boston College; Chestnut Hill, MA.

**Contents of this file**

## Introduction

This Supporting Information provides additional details on alternative products and calculation procedures.

Dataset: ARCONNECT_v2 | Period: 2000–2019

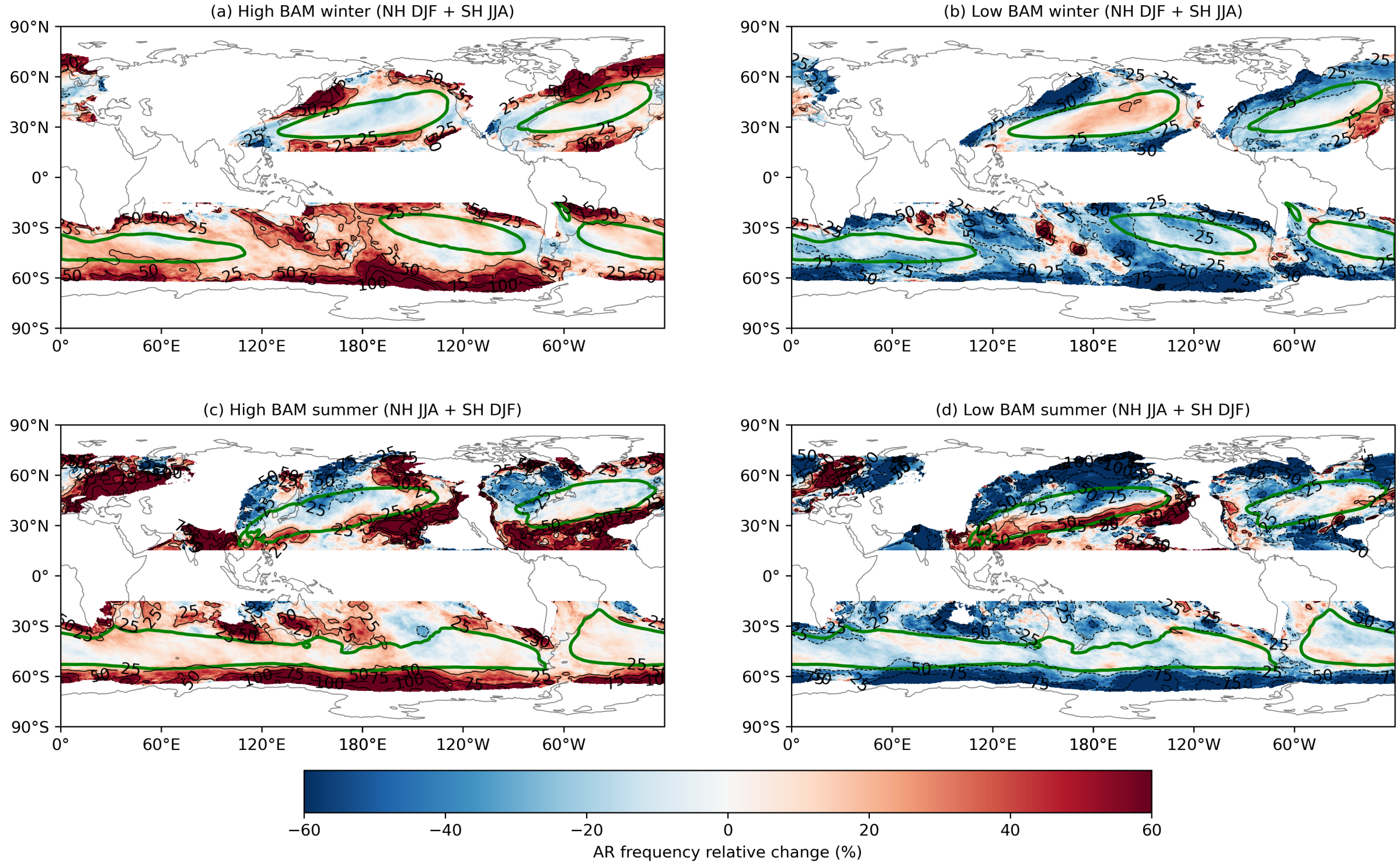

**Figure S1. Alternative AR products.** BAM-state composite of relative changes in AR occurrence based on ARCONNECT_v2 product during 2000–2019: (a) high-BAM winter, (b) low-BAM winter, (c) high-BAM summer, and (d) low-BAM summer states. Winter and summer are defined locally: DJF and JJA, respectively, in the NH, and JJA and DJF, respectively, in the SH. Shading shows the difference between composite and climatological AR occurrence frequencies normalized by the corresponding seasonal climatology; black contours are plotted at 25% intervals. Thick green contours denote a climatological AR occurrence frequency of 30% for the corresponding season. The tropical region from 15°S to 15°N and grid points with climatology AR occurrence frequencies below 2% are masked.

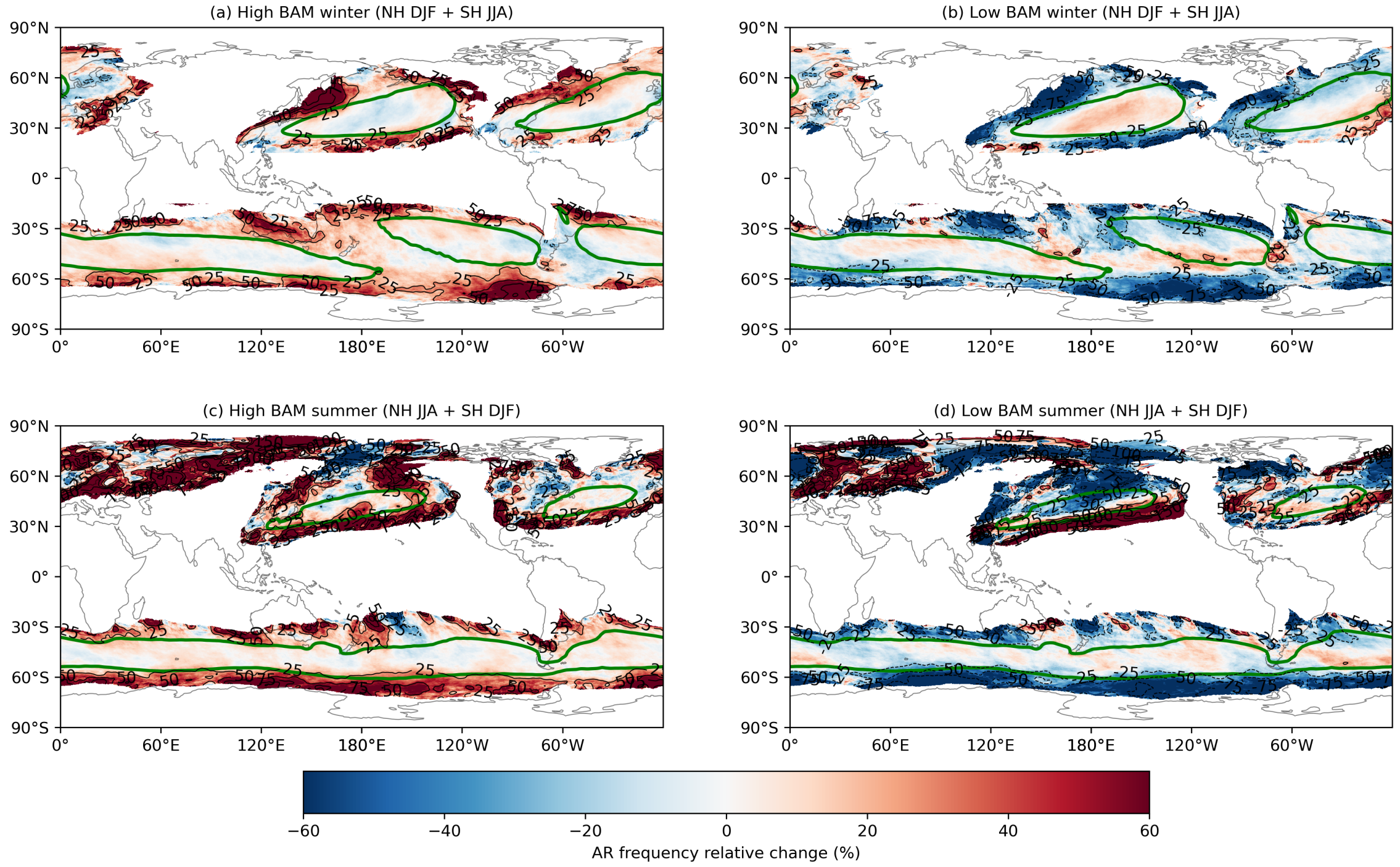

**Figure S2.** As in Figure S1, but for Lora_v2 product during 2000–2019.

Dataset: Mundhenk_v3 | Period: 1980–2019

(a) High BAM winter (NH DJF + SH JJA)

(b) Low BAM winter (NH DJF + SH JJA)

(c) High BAM summer (NH JJA + SH DJF)

(d) Low BAM summer (NH JJA + SH DJF)

AR frequency relative change (%)

**Figure S3.** As in Figure S1, but for Mundhenk_v3 product during 1980–2019.

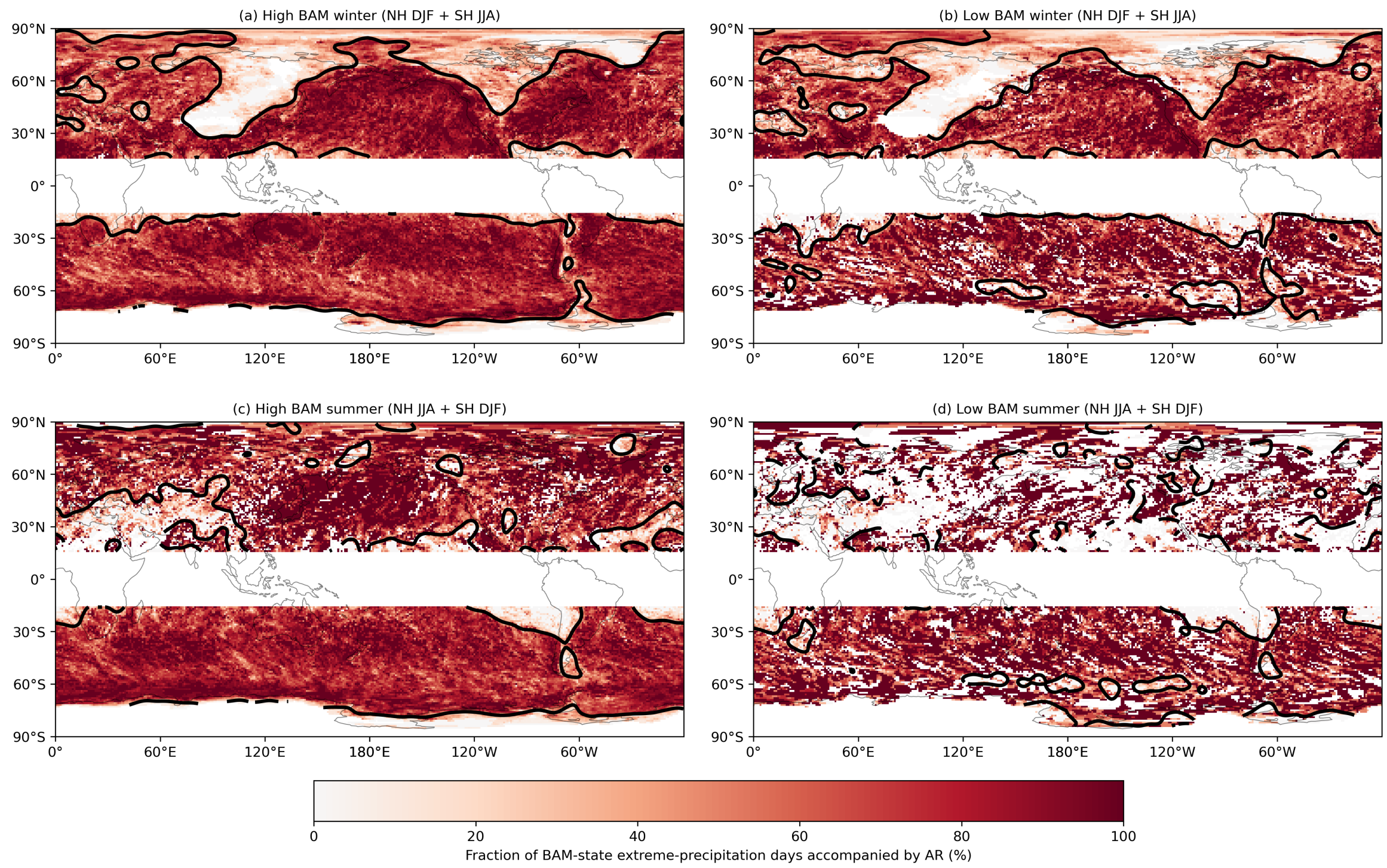

**Figure S4.** Fraction of extreme-precipitation days during each BAM state associated with ARs. Thick black contours show the 50% level.